\documentclass[aps, prx, reprint, superscriptaddress, showpacs, nofootinbib, longbibliography, floatfix]{revtex4-2}

\usepackage{graphicx}% Include figure files
\usepackage{dcolumn}% Align table columns on decimal point
\usepackage{bm}% bold math
\usepackage{amsmath,amsthm, amssymb}
\usepackage{graphicx}
\usepackage{comment}
\usepackage{parskip}
\newcommand{\clifford}{\mathrm{Cl}}
\newcommand{\pclifford}{\mathrm{Proj\: Cl}}
\newcommand{\sym}{\mathrm{Sp}}
\newcommand{\ord}{\mathrm{ord}}
\newcommand{\id}{\mathrm{Id}}
\usepackage{color}

\begin{document}

\preprint{APS/123-QED}

\title{No-Go Theorem on Fault Tolerant Gadgets for Multiple Logical Qubits}% Force line breaks with 

\author{Aranya Chakraborty}
 \altaffiliation{aranyac@umd.edu}%Lines break automatically or can be forced with 
\author{Daniel Gottesman}%
\altaffiliation{dgottesm@umd.edu}
\affiliation{%
Department of Physics, University of Maryland, College Park ; QUICS
}%

\date{\today}% It is always \today, today,

\begin{abstract}
    Identifying stabilizer codes that admit fault-tolerant implementations of the full logical Clifford group would significantly advance fault-tolerant quantum computation, as such implementations prevent uncontrolled error propagation. Motivated by this goal, we study several classes of fault-tolerant gadget constructions consisting of Clifford gates acting on the physical qubits, including transversal gadgets, code automorphisms, and fold-transversal gadgets. While stabilizer codes encoding a single logical qubit — most notably the $[[7,1,3]]$ Steane code — are known to admit transversal implementations of the full logical Clifford group, no analogous examples are known for codes encoding multiple logical qubits. In this work, we prove a no-go theorem establishing that no stabilizer code admits a fully transversal implementation of the Clifford group on more than one logical qubit. We further strengthen this result by showing that fold-transversal implementations of the full logical Clifford group are impossible for stabilizer codes encoding more than two logical qubits. More generally, we introduce the notion of $k$-fold transversal gadgets and prove that implementing the full Clifford group on $k$ logical qubits requires at least $k$-fold transversal gadgets at the physical level. In addition, we analyze code-automorphism based constructions and demonstrate that they also fail to realize the full Clifford group on multiple logical qubits for any stabilizer code. Together, these results place fundamental constraints on fault-tolerant Clifford gadget design and show that stabilizer codes supporting the full logical Clifford group on multiple logical qubits via these architectures do not exist. Since the Clifford group is a core component of universal gate sets, our findings imply that quantum computing with codes encoding multiple logical qubits within a single code block necessarily entails more complex constructions for fault tolerance.
\end{abstract}

\maketitle

%\tableofcontents

\section{Introduction}

Quantum computers leverage the laws of quantum mechanics in order to solve a variety of problems in physics and chemistry far beyond the capabilities of a classical computer \cite{majidy2024building}. However, working on quantum computers also leaves us vulnerable to a much wider range of noise than seen on a classical computer. Thus, one of the cornerstones in realizing the \textit{quantum advantage} is the ability to encode the logical information and protect it from errors. The most widely used and reliable way to protect quantum information is by using the class of stabilizer codes \cite{gottesman1998theory}, where we encode the logical information onto a larger number of highly entangled physical qubits. This allows us to protect the encoded information from the inherent noise in a quantum system as long as the error is below a certain threshold \cite{aharonov1997fault, knill1998resilient}. Recently,  significant strides have been made towards demonstrating practical error correction in a variety of qubit architectures \cite{reichardt2024demonstration,reed2012realization,paetznick2024demonstration, google2025quantum}. However, in order to do meaningful computation, in addition to encoding the information we need a way of performing operations on the encoded logical information. In particular, a class of logical operations that are of interest is the Clifford group ($\clifford_n$) due to its significance in quantum error correction and simulation \cite{gottesman1998heisenberg}. More importantly, the Clifford group also has an important role in the construction of universal gate sets \cite{kliuchnikov2012fast} as any non-Clifford gate in addition to the full Clifford group gives us the universal gate set \cite{dawson2005solovay, aharonov2003simple}.

At the same time we need a way to perform the desired logical operations in a \textit{fault tolerant} manner, that is without spreading errors on to multiple qubits, thereby preventing uncontrolled error propagation. A natural approach is to employ transversal gates, which are composed of single-qubit Clifford operations acting independently on each physical qubit. Because transversal gates do not involve multi-qubit interactions, they are inherently fault tolerant and suppress the propagation of errors during implementation. Moreover, the product of two transversal gates is itself transversal; consequently, any logical operation realizable through transversal gates can be implemented with constant circuit depth.

Even though no stabilizer code admits a transversal implementation of the universal gate set \cite{eastin2009restrictions, zeng2011transversality}, there exist various methods to introduce non-Clifford gates including magic state injection \cite{knill2004fault, bravyi2005universal} and error correction \cite{paetznick2013universal}. These methods, combined with stabilizer codes having a transversal implementation of the full logical Clifford group, provide the most realistic method to obtain fault tolerant universal quantum computation in the near term. Several quantum codes encoding a single logical qubit and permitting transversal implementation of the entire logical Clifford group have been identified, with the $[[7,1,3]]$ Steane code \cite{steane1996error} being the smallest such example. More recently, attention has shifted toward quantum codes that encode multiple logical qubits and support addressable transversal gates \cite{guyot2025addressability, he2025quantum}, which allow us to interact with each logical qubit independently. Such codes not only provide a better rate of encoding and reduce qubit overhead but also allow more flexible code switching and gauge fixing frameworks \cite{paetznick2013universal, bombin2015gauge} and enable more efficient implementations of non-Clifford operations \cite{bravyi2012magic,bravyi2013classification}. This could allow us to circumvent Eastin-Knill restrictions and expand the accessible fault-tolerant gate sets.

However, the number of logical actions that we can perform while limiting to transversal gadgets at the physical level is usually small for general stabilizer codes and indeed no stabilizer code has been found that supports the transversal implementation of the full logical Clifford group for multiple qubits. Hence, there exists a need to expand the notion of transversality and there exist two leading ways to achieve this. The first is by using code automorphisms where we allow for qubit permutations in addition to transversal Clifford gates at the physical level \cite{calderbank1998quantum, sayginel2024fault, berthusen2025automorphism, koh2026entangling}.  The second is by using fold-transversal gates \cite{sahay2025fold, breuckmann2024fold, eberhardt2024logical, moussa2016transversal, quintavalle2023partitioning} which use multi-qubit Clifford gates within a single physical code block, which spread errors but can implement a greater number of logical actions.

In this paper, we establish a no-go theorem that limits the performance of both transversal and fold transversal Clifford gadgets, proving that no stabilizer code has a Clifford transversal implementation of the full Clifford group on multiple logical qubits or a Clifford fold transversal implementation of the full Clifford group on more than two logical qubits. 

The theorem can be generalized for $[[n,k,d]]$ codes by introducing $k$-fold transversal gadgets and proving that in order to implement the full logical Clifford gate on $k$ logical qubits we need to implement at least $k$-fold transversal Clifford gadgets at the physical level. In the worst case, a $k$-fold transversal gadget can propagate a single qubit error on to $k$ physical qubits and hence becomes decreasingly fault tolerant as $k$ increases.  An important example of $k$-fold transversal gates are transversal gates between $k$ blocks of a code.  Our theorem shows that even by using multiple blocks, Clifford transversal gates cannot achieve the full logical Clifford group unless each block has only a single logical qubit.

We further prove a constraint on using code automorphisms as physical gadgets and show that they can never implement the full logical Clifford group for any stabilizer code with multiple logical qubits. Thus the theorems presented in this paper establish a clear constraint on high rate codes, limiting the number of logical actions which can be implemented using the fault tolerant Clifford gadgets considered in this paper. As a result it becomes important to explore other pathways to universal fault tolerant computation including code switching \cite{butt2024fault, daguerre2025experimental, paetznick2013universal, tan2025single}, flag fault tolerance \cite{chao2018quantum, anker2025universal, berthusen2025adaptive}, lattice surgery \cite{horsman2012surface, fowler2018low} etc.

The rest of the paper is organized as follows. In Section \ref{Pre} we go over some preliminaries which will be required for the rest of the paper. We then define the central theorems of our paper in Section \ref{main} and provide a proof for the theorems in Section \ref{proof}. We then end our work with a discussion on the impact of the theorems presented and future directions which can be pursued in Section \ref{dis}.

\section{Preliminaries} \label{Pre}

\subsection{Stabilizer Codes} 

The stabilizer formalism provides a compact description of a large class of quantum codes and operations \cite{gottesman1998theory}.  
A \([[n,k,d]]\) stabilizer code encodes \(k\) logical qubits into \(n\) physical qubits with distance \(d\).  
It is defined by an abelian subgroup \(\mathcal{S} \subset \mathcal{P}_n\) of the \(n\)-qubit Pauli group
\begin{equation}
    \mathcal{P}_n = \langle i I, X_j, Z_j : j=1,\dots,n \rangle ,
\end{equation}
where each element of \(\mathcal{S}\) fixes the codespace.  
The codespace is therefore the joint \(+1\) eigenspace of all stabilizer generators.

\textit{a. Binary symplectic representation : }
Every Pauli operator on \(n\) qubits can be written uniquely (up to phase) as
\begin{equation}
    P(\mathbf{a},\mathbf{b}) =  
    \bigotimes_{j=1}^n X_j^{a_j} Z_j^{b_j},
\end{equation}
with binary vectors \(\mathbf{a},\mathbf{b} \in \mathbb{F}_2^n\).  
This identifies each Pauli operator with a binary vector
\begin{equation}
    (\mathbf{a}\,|\,\mathbf{b}) \;\in\; \mathbb{F}_2^{2n}.
\end{equation}
The commutation relation between two Pauli operators \(P(\mathbf{a},\mathbf{b})\) and \(P(\mathbf{a}',\mathbf{b}')\) is determined by the \emph{symplectic inner product}
\begin{equation}
\begin{split}
        \langle (\mathbf{a},\mathbf{b}), (\mathbf{a}',\mathbf{b}') \rangle &= (\mathbf{a},\mathbf{b})^T  J (\mathbf{a}',\mathbf{b}') \\
    &= \mathbf{a}\cdot\mathbf{b}' + \mathbf{b}\cdot\mathbf{a}' \;\;\; (\text{mod }2).
\end{split}
\end{equation}

where, $J = \begin{bmatrix}
    0 & \id_n  \\ \id_n & 0
\end{bmatrix}$.

Two Paulis commute if their symplectic inner product vanishes. The stabilizer \(\mathcal{S}\) can be represented as the row space of a binary \((n-k)\times 2n\) matrix
\begin{equation}
    S = \begin{bmatrix}
        \mathbf{a}_1 & \mathbf{b}_1 \\
        \vdots & \vdots \\
        \mathbf{a}_{n-k} & \mathbf{b}_{n-k}
    \end{bmatrix},
\end{equation}
with rows encoding the stabilizer generators.

\subsection{Clifford Group}
The Clifford group $\clifford_n$ is defined as the normalizer of $\mathcal{P}_n$, i.e.
\begin{equation}
    U \in \clifford_n \quad \iff \quad U P U^\dagger \in \mathcal{P}_n \;\; \forall P \in \mathcal{P}_n.
\end{equation}

We define the projective Clifford group as the Clifford group mod phases and the Paulis, i.e. $
\pclifford_n \;:=\; \clifford_n \big/ \big( U(1),\mathcal{P}_n \big)$. The projective Clifford group is isomorphic to the classical symplectic group such that for each $U \in \pclifford_n$ there exists a $M_U \in \sym(2n,2)$ such that,
\begin{equation}
    U^\dag P(\mathbf{a},\mathbf{b})U = P(\mathbf{a'},\mathbf{b'})  \:\: \Leftrightarrow \:\: \mathrm{ M_U \cdot (\mathbf{a},\mathbf{b})  = (\mathbf{a'},\mathbf{b'})}.
\end{equation}

Thus, each operator in the projective Clifford group can be represented as $2n \times 2n$ binary matrix, acting linearly on the binary representation of the Pauli operators. Since Clifford operators are unitaries, they must preserve the commutation relation between Pauli operators and this constraint is equivalent to their binary representation satisfying the symplectic condition. 
\begin{equation}
    M_U^T J M_U = J,
\end{equation}

For instance, on a single qubit:
\begin{align}
    H: &\quad 
    \frac{1}{\sqrt{2}}\begin{bmatrix}1 & 1 \\ 1 & -1 \end{bmatrix}
    \mapsto M_H = 
    \begin{bmatrix}0 & 1 \\ 1 & 0 \end{bmatrix}, \\
    S: &\quad
    \begin{bmatrix}1 & 0 \\ 0 & i \end{bmatrix}
    \mapsto M_S = 
    \begin{bmatrix}1 & 0 \\ 1 & 1 \end{bmatrix}.
\end{align}

The \textbf{order} of a Clifford operator $U \in \clifford_n$ is defined as the smallest integer $r$ such that 
\begin{equation}
    U^r = \id_n.
\end{equation}

Since the Clifford group is a finite group, all Clifford operators have a finite order. For example, the single qubit projective Clifford group only has elements having order : 1 (Eg. $\id$) , 2 (Eg. $H \: \mathrm{and}\: S$) and  3 (Eg. $HS$).

\subsection{Field Theory}

A field $\mathbb{F}$ is an algebraic structure where mathematical operations such as addition, subtraction, multiplication etc. are well defined. Since we are working with qubits, a field of particular importance to us is the binary field $\mathbb{F}_2 = \{0,1\}$. Finite fields having sizes which are powers of primes can be constructed using field extensions: for any finite field $\mathbb{F}_p$ of size $p$ and for each prime power $q = p^m$, there exists a unique field $\mathbb{F}_q$ which can be constructed as 
\begin{equation}
    \mathbb{F}_q \cong \mathbb{F}_p[x] / (f(x)),
\end{equation}
where $\mathbb{F}_p[x]$ is the ring of polynomials with coefficients in the field $\mathbb{F}_p$ and $f(x)$ is an irreducible polynomial of degree $m$.

Let $\alpha = x \mod f(x)$. $\alpha$ is then called the residue class of $x$ and every element of $\mathbb{F}_q$ can be written as 
\begin{equation}
    a_0 + a_1 \alpha + \cdots + a_{m-1}\alpha^{m-1} ,  \quad a_i \in \mathbb{F}_p.
\end{equation}

An irreducible polynomial is called \textit{primitive} if $\alpha$ has multiplicative order $p^m - 1$; in this case, $\alpha$ generates the multiplicative group $\mathbb{F}_q^\times = \mathbb{F}_q \backslash \{0\}$ that is, 
\begin{equation*}
    \mathbb{F}_q^\times = \langle\alpha\rangle = \{\alpha^k : 1 \leq k \leq p^m - 1\}.
\end{equation*}

\textbf{Example.} Let us define $\mathbb{F}_9 = \mathbb{F}_{3^2}$ using the primitive polynomial 
\begin{equation}
    f(x) = x^2 + x + 2 \in \mathbb{F}_3[x],
\end{equation}
and let $\alpha = x \mod f(x)$.  Then $\alpha$ satisfies $\alpha^2  = 2\alpha + 1$. Every element of $\mathbb{F}_9$ can therefore be uniquely written as
\begin{equation}
    a_0 + a_1 \alpha,  \quad a_i \in \mathbb{F}_3.
\end{equation}

Then, the multiplicative group $\mathbb{F}_9^\times$ can be generated by successive powers of $\alpha$ such that
\begin{equation}
\begin{split}
    &\alpha^1  = \alpha \\
    &\alpha^2 = 2\alpha+1 \\
    &\alpha^3 = 2\alpha^2+ \alpha = 2\alpha+2 \\
    &\alpha^4 = 2\alpha^2+ 2\alpha = 2 \\
    &\alpha^5 = 2\alpha \\
    &\alpha^6 = 2\alpha^2 = \alpha +2 \\
    &\alpha^7 = \alpha^2 + 2\alpha = \alpha+1\\ 
    &\alpha^8 = \alpha^2 + \alpha = 1.
\end{split}
\end{equation} \qed

\subsection{Fault Tolerance}

In order to do computation using stabilizer codes, we use physical \textit{gadgets} $\in \mathrm{SU}(n)$ which act on the physical qubits and perform some \textit{logical action} $\in \mathrm{SU}(k)$ on the logical qubits. In this paper, we limit our attention to the physical Clifford group and only consider gadgets $\in \clifford_n$. In order for a physical operation ($U$) to be a gadget it must preserve the code space i.e. map codewords to other codewords. Equivalently, for stabilizer codes, the above condition is equivalent to 
\begin{equation}
    U^\dag S U \in \mathcal{S} \quad \forall\: S \in \mathcal{S}
\end{equation}

However, in order to prevent fault propagation we need the physical gadgets to be \textit{fault tolerant}. One of the best way to ensure fault tolerance is to make the physical gadgets local and minimize multi-qubit operators. Inherently fault tolerant gadgets, known as transversal gadgets, act independently on each individual physical qubit,
\begin{equation}
    U_{\text{trans}} = \bigotimes_{i=1}^n U_i 
\end{equation}
and prevent fault propagation within a code block. We further limit each $U_i$ to be in the single qubit Clifford group and call these gadgets, Clifford transversal gadgets.

The term transversal is also used for gadgets between $k$ code blocks but now each $U_i$ acts on qubit number $i$ within each of the $k$ blocks.  Thus, each $U_i$ acts only one qubit within a block.  In this paper, we focus on transversal gadgets on single blocks, but multiple-block transversal gadgets can be seen as an example of $k$-fold transversal gadgets, which we introduce in section~\ref{main}.

The idea of transversal gates has been expanded recently to include some two qubit operations within a single code block.  These are known as fold transversal gadgets. They are constructed with respect to a chosen ZX duality $\tau$.  A unitary $U$ is considered to be fold transversal if it is supported on $\{i, \tau(i)\}$ for $i = 1, 2, \cdots, n$ \cite{berthusen2025simple}. We further restrict $U \in \clifford_n$ and call them Clifford fold transversal gadgets. These gadgets exploit the geometrical properties of the code and only propagate single qubit errors to two qubit errors which have independent syndromes. This allows us to identify and correct these two qubit errors which show up most commonly during implementation even if these codes can't correct any arbitrary two qubit errors.

The idea of transversal gates can also be relaxed on certain qubit architectures such as Rydberg atoms \cite{xiao2024effective}, ions \cite{kielpinski2002architecture} etc. In these architectures, SWAP gates can also be made fault tolerant since qubit relabeling can be done in software, enabling us to construct fault tolerant gadgets made up of a Clifford transversal gadget followed by a permutation of physical qubits. Such gadgets are known as code automorphisms.

In order to prevent the spread of errors it is important to enforce as strict a notion of fault tolerance as possible. In terms of their fault tolerant properties, that is how well they are able to limit the spread of errors, the gadgets can be ranked as follows
\begin{equation*}
\substack{\text{Transversal}  \\\text{Gadgets}}
\;\geq\;
\substack{\text{Code}  \\\text{Automorphisms}}
\;>\;
\substack{\text{Fold Transversal}  \\\text{Gadgets}}.
\end{equation*}

However, sacrificing on fault tolerance could allow us more flexibility in terms of the logical actions available to us. 

\section{Central Theorem} \label{main}

Transversal gadgets, though a very desirable fault tolerant method to construct physical gadgets, are very limited in their ability to perform logical actions for most stabilizer codes. Our first theorem provides a fundamental constraint on the logical actions that can be implemented by using transversal Clifford gadgets at the physical level. We limit ourselves only to the logical Clifford group since it is an important ingredient to achieve universal quantum computation. We then show that for stabilizer codes having more than one logical qubit, it is impossible to perform the whole logical Clifford group using just transversal Clifford gadgets. In order to generalize it to stabilizer codes having multiple logical qubits we introduce the notion of $k$-fold transversal gadgets (similar to the notion of transversality introduced in \cite{jochym2018disjointness}). This notion generalizes the concept of fold transversality and under this convention, transversal gates are $1$-fold transversal gadgets and fold transversal gates under a fixed pairing of physical qubits are an example of $2$-fold transversal gadgets.

A set of gadgets ($G$) is $k$-fold transversal if there exists a disjoint partition of physical qubits: 

\begin{equation}
    \{s_i\}_{i=1}^m \quad : \quad \bigcup\limits_{i=1}^m s_i  = \{1,2,...,n\},
\end{equation}
such that  
 \begin{equation}
     |s_i| \leq k \quad \forall \: i,
 \end{equation}
and every $U \in G$ can be expressed as a tensor product
\begin{equation}
    U = \otimes_{i=1}^m U_{s_i},
\end{equation}
where $U_{s_i}$ acts only on the qubits in $s_i$. In this work, we restrict $U_{s_i} \in \clifford_{|s_i|}$ and call these gadgets Clifford k-fold transversal gadgets.

Under this definition, k-fold transversal gadgets share a fundamental property with both transversal gadgets and fold-transversal gadgets. If U and V are k-fold transversal with respect to a given partition, then their product UV is also k-fold transversal for the same partition. Consequently, the collection of k-fold transversal gadgets associated with a fixed partition forms a group. It follows that any logical operation realizable through k-fold transversal gadgets can be implemented with constant circuit depth.

This is a significant property, since for every stabilizer code, fold-transversal gates over arbitrary partitions always permit implementation of the full logical Clifford group. This is because $\{H,S,CNOT\}$, which is a 2 qubit gate set, allows us to implement arbitrary physical Clifford circuit and thus all logical Clifford gates. However, these gates no longer form a group structure themselves and, as a result, implementing an arbitrary logical Clifford operation would generally require non-constant circuit depth.

We now present the first theorem.

\textbf{Theorem 1.} In order to implement the full Clifford group on stabilizer codes with $k$ logical qubits, we need at least Clifford $k$-fold transversal gadgets acting on the physical qubits.

\textbf{Corollary.} There exists no stabilizer code which permits a Clifford transversal implementation of the full Clifford group on multiple logical qubits or a Clifford fold transversal implementation of the full Clifford group on more than two logical qubits.

The corollary follows from the definition of the $k$-fold transversal gates. A simple example to illustrate Theorem 1 would be to consider $k$ blocks of the $[[7,1,3]]$ code. The full logical Clifford group can be performed for this code by using transversal gadgets in each code block and transversal CNOT gadgets between code blocks. However, we can also think of the aggregate code as a single $[[7k, k, 3]]$ code. Since the CNOT gates now act within a single code block, they are no longer transversal and are in fact $k$-fold transversal (as they act on one qubit from each block). Thus, this is a straightforward example of a stabilizer code having the full Clifford group on $k$ logical qubits implemented using Clifford $k$-fold transversal gadgets, which also shows our theorem is tight. An important observation is that, if one instead considers k blocks of a CSS code encoding two logical qubits, yielding an $[[nk,2k,d]]$ code, then even in the presence of transversal inter-block CNOT gates, one can realize at most the full logical Clifford group on only k logical qubits.

We present another theorem that constrains the logical actions of code automorphisms at the physical level. Code automorphisms, similar to $k$-fold transversal gadgets, form a group therefore any logical operation realizable through code automorphisms can be implemented with constant circuit depth.

\textbf{Theorem 2.} There exists no stabilizer code that admits the full Clifford group on multiple logical qubits implemented using code automorphisms.

The above two theorems provide a fundamental constraint on the logical actions available for a stabilizer code using fault tolerant Clifford gadgets such as fold transversal gates and code automorphisms. This shows that in order to do the full Clifford group on multiple logical qubits in a fault tolerant manner, more complex constructions would be required.

\section{Proof of the Central Theorems} \label{proof}

\subsection{Proof of Theorem 1}

In this section, we provide a proof for Theorem 1 stated in Section \ref{main}. We first show a correspondence between the order of the physical gadget acting on the physical qubits and the logical operation implemented by it.

\textbf{Lemma 1 [Order Lemma]:} The order of the physical gadget must be a multiple of the order of the logical action induced by it.

\textbf{Proof.} Let $U \in \mathrm{Cl}_n$ be the physical gadget on the physical qubits and $\bar{U} \in \mathrm{Cl}_k$ be its logical action and let $\mathrm{ord}(U) = r$. Then, $U^r$ acts trivially on the physical qubits and would by definition induce a trivial logical action i.e.

\begin{equation}
    U^r = \mathrm{Id_n} \quad \Rightarrow \quad \overline{U}^r = \overline{\mathrm{Id_k}}.
\end{equation}

Hence, the logical action induced by the physical gadget $U$ must have an order that is a factor of $\mathrm{ord}(U)$. \qed

\textbf{Corollary.} If the order of the physical gadget is prime, then the induced logical action either has the same order or acts trivially on the logical qubits.

The above lemma is valid for any physical gadget $U \in \mathrm{SU(n)}$ but we will be restricting our attention to physical gadgets belonging to a subset of the full unitary group, the Clifford group ($\clifford_n$). Physical Clifford gadgets always induce a Clifford logical action on a stabilizer code
\begin{equation*}
    U \in \clifford_n \Rightarrow \overline{U} \in \clifford_k.
\end{equation*}

An important thing to note is that elements in the Clifford group can in principle have any order by adding a phase term. However, since a phase as well as a Pauli gadget induces a trivial logical action we would be ignoring their effect and restricting ourself to physical gadgets in the projective Clifford group.

The key idea of the proof is to show that there exists a unitary operation in $\clifford_k$ of prime order that cannot be realized within $\clifford_{k-1}$. Consequently, to implement this Clifford operation on the logical qubits, one must realize a physical gadget belonging to $\clifford_k$, corresponding to a $k$-fold transversal gate. We show this with help of the following lemmas.

\textbf{Lemma 2 [Zsigmondy's Theorem]:} There exists a prime $p$ which divides $2^{2k} - 1$ and doesn't divide $2^{2i} - 1$ for all $i < k$.

\textbf{Proof.} The above lemma is an application of Zsigmondy's theorem which was first presented in \cite{zsigmondy1892theorie} (For a more abridged proof see \cite{michels2014zsigmondy}). The theorem states that for any $a,b$ which are positive co-prime integers, and any integer $n \geq 1$, there exists a prime (also known as primitive prime divisor) number $p$ that divides $a^{n}-b^{n}$ and does not divide $a^{i}-b^{i}$ for any positive integer $i < n$. 

Lemma 2 is a specific case of the above theorem with $a = 2$ and $b = 1$ (This simplification is also known as Bang's theorem and was proved earlier in \cite{bang1886taltheoretiske}). We further simplify by only considering even exponents i.e. $n = 2k$ for any integer $k \geq 1$. An important thing to note is that Bang's theorem has an exception for $n = 6$, where we get $2^6 - 1 = 63 = 7 \times 3^2$ and the two primes $3$ and $7$ appear previously for $n=2$ and $n=3$ respectively and hence we don't get a primitive prime divisor. However, as we are only considering even exponents, the above exception is no longer valid and $7$ becomes a ``primitive" prime divisor and our lemma can be applied without exceptions. \qed

From Lemma 2, we know there exists a prime $p$ that divides $2^{2k} - 1$ but no smaller term of that form. Such a prime must divide either $2^k + 1$ or $2^k - 1$. We now show that there exists a Clifford operator in $\clifford_k$ that has order $2^k + 1$ and $2^k - 1$.

\textbf{Lemma 3 [Existence Lemma]:} There exists a $V, W \in \clifford_k$ such that $\mathrm{ord(V) = 2^k - 1}$ and $\mathrm{ord(W) = 2^k + 1}$.

\textbf{Proof.} As shown in section \ref{Pre}, each gate in the projective Clifford group $\clifford_k$ can be written as a matrix in the classical Symplectic group $\sym(2k,2)$. Hence, it is enough to show that there exists a classical matrix, $M$ of dimension $2k \times 2k$ which satisfies that symplectic condition, $M^T J M = J$, and has the required order.

\textbf{(a.)} We first show how to construct an operator $V \in \sym(2k, 2)$ such that $\mathrm{ord(V) = 2^k - 1}$. 
\begin{enumerate}
    \item Select a primitive polynomial
    \begin{equation*}
        p(x)=x^k + a_{k-1}x^{\,k-1} + \cdots + a_1 x + a_0 \in \mathbb{F}_2[x],
    \end{equation*}
    
    of degree $k$ and construct the field extension $\mathbb{F}_{2^k}$ 
    \begin{equation}
        \mathbb{F}_{2^k} \cong \mathbb{F}_2[x]/(p(x)).
    \end{equation}
    Let $\alpha$ denote the residue class of $x$.  
    Since $p(x)$ is primitive, $\alpha$ has multiplicative order $2^k - 1$.

    \item Construct the $k \times k$ dimensional companion matrix of $p(x)$ given by
    \begin{equation}
        C_p=
    \begin{bmatrix}
        0 & 0 & \cdots & 0 & a_0 \\
        1 & 0 & \cdots & 0 & a_1 \\
        0 & 1 & \cdots & 0 & a_2 \\
        \vdots & \vdots & \ddots & \vdots & \vdots \\
        0 & 0 & \cdots & 1 & a_{k-1}
    \end{bmatrix}.
    \end{equation}
    With respect to the basis $\{1,\alpha,\ldots,\alpha^{k-1}\}$, this matrix represents
    multiplication by $\alpha$ on $\mathbb{F}_{2^k}$.  
    Because $\alpha$ has multiplicative order $2^k - 1$, the matrix $C_p$ also has multiplicative order $2^k - 1$.

    \item Using $C_p$ we construct a symplectic matrix $\in \sym(2k,2)$ 
    \begin{equation}
        V = \begin{bmatrix}
            C_p & \mathbf{0}_{k \times k} \\
            \mathbf{0}_{k \times k} & (C^{-1}_p)^T
        \end{bmatrix}.
    \end{equation}

    The matrix $V$ always satisfies the symplectic condition, $V^T J V = J$ and moreover
    \begin{equation}
        V^r = \begin{bmatrix}
            C^r_p & \mathbf{0}_{k \times k} \\
            \mathbf{0}_{k \times k} & (C^{-r}_p)^T
        \end{bmatrix}.
    \end{equation}

    Hence, $V$ has the same order as $C_p$, that is $\mathrm{ord(V) = 2^k - 1}$.
\end{enumerate}

\textbf{(b.)} We then show how to construct an operator $W \in \sym(2k, 2)$ such that $\mathrm{ord(W) = 2^k + 1}$.

\begin{enumerate}
    \item Select a monic primitive polynomial $p(x)$ of degree $2k$ and construct
    \begin{equation}
        \mathbb{F}_{2^{2k}} \cong \mathbb{F}_2[x]/(p(x)).
    \end{equation}
    On $\mathbb{F}_{2^{2k}}$ we use the standard bilinear form (used in Equation 8.41 in \cite{gottesman2024surviving})
    \begin{equation}
        B(a,b) = \operatorname{Tr}_{\mathbb{F}_{2^{2k}}/\mathbb{F}_2}
        \left( \theta \cdot (a b^{2^k} - a^{2^k} b)\right),
    \end{equation}
    which is known to be alternating and can be made non-degenerate by choosing a $\theta \in \mathbb{F}_{2^{2k}}$.
    Therefore there exists a basis of $\mathbb{F}_{2^{2k}}$ over $\mathbb{F}_2$ in which
    \begin{equation}
        B(a,b) = a^T J b,
    \end{equation}
    where $J=\begin{bmatrix} 0 & \id_k \\ \id_k & 0 \end{bmatrix}$.

    \item Let $\alpha$ be the residue class of $x$ in $\mathbb{F}_{2^{2k}}$.  
    Since $p(x)$ is primitive, $\alpha$ has multiplicative order $2^{2k}-1$.
    Define
    \begin{equation}
        t = \alpha^{\,2^k - 1}.
    \end{equation}

    The element $t$ has multiplicative order $2^k + 1$.
    Let $W$ be the linear transformation on $\mathbb{F}_{2^{2k}}$ defined by
    \begin{equation}
        Wa = t \cdot a.        
    \end{equation}

    Thus
    \begin{equation}
        \operatorname{ord}(W) = 2^k + 1.        
    \end{equation}

    \item The matrix $W$ preserves the symplectic form $B$ that is
    \begin{equation*}
    \begin{split}
        B(Wa, Wb) &= \operatorname{Tr}_{\mathbb{F}_{2^{2k}}/\mathbb{F}_2}
        \left( \theta \cdot (ta \cdot (tb)^{2^k} - (ta)^{2^k} \cdot tb)\right)  \\
        &= \operatorname{Tr}_{\mathbb{F}_{2^{2k}}/\mathbb{F}_2} \big(\theta \cdot t^{2^k + 1}(a b^{2^k} - a^{2^k} b)\big) \\
        & = \operatorname{Tr}_{\mathbb{F}_{2^{2k}}/\mathbb{F}_2}
        \left( \theta \cdot (a b^{2^k} - a^{2^k} b)\right) \\
        &= B(a,b) \quad \quad \forall \: a,b.
    \end{split}
    \end{equation*}

Hence, 
    \begin{equation}
    \begin{split}
            a^T J b &= (Wa)^T J (Wb) \\
            &  =a^T (W^T JW) b  \quad\quad \forall a,b.
    \end{split}
    \end{equation}

    Hence, $W^T J W = J$. We construct an explicit example to illustrate Lemma 3 in Appendix \ref{app1}.\qed
\end{enumerate}

Since we know that $p$ divides either $2^k + 1$ or $2^k - 1$ then by Lemma 3, either $U = V^{\frac{2^k - 1}{p}}$ or $U = W^{\frac{2^k + 1}{p}}$ will be a valid Clifford operator $\in \clifford_k$ such that $\ord(U) = p$. We now show that such an operator can't exist in $\clifford_{k-1}$.

\textbf{Lemma 4 [Uniqueness Lemma]:} The order of any $U \in \pclifford_k$ must divide 
\begin{equation}
    2^{k^2} \prod_{i = 0}^k (2^{2i} - 1).
\end{equation}

\textbf{Proof.} Weyl showed in \cite{weyl1946classical} that the order of the symplectic group over the binary
field is
\begin{equation}
        |\mathrm{Sp}(2k,2)| 
    = 2^{k^2}\prod_{i=0}^{k} (2^{2i}-1).
\end{equation}

As shown in Section \ref{Pre}, the projective Clifford group is isomorphic to the classical symplectic group and thus has the same size. Let $U \in \clifford_k$ have order $r$. The cyclic subgroup it generates,
\begin{equation}
    \langle U \rangle = \{ I, U, U^2, \ldots, U^{r-1} \},
\end{equation}
has size $r$. By Lagrange's Theorem, the order of any subgroup of a finite group divides the order of
the group itself. Thus the order of any element of the Clifford group must divide $2^{k^2}\prod_{i=0}^{k}(2^{2i}-1)$. \qed

Since Lemma 2 guarantees that the prime $p$ doesn't divide $(2^{2i} - 1)$ for any $i < k$ using Lemma 4, we show that an operator having an order $p$ can never exist in $\pclifford_{k-1}$. Since appending a Pauli can only multiply the order of an operator by a power of 2, there can not exist an operator of order $p$ even in $\clifford_{k-1}$ where p is the primitive prime divisor.

The above Lemmas combine to show that there exists a operator in $U \in \clifford_k$ which has an order $p$ where $p$ is the primitive prime divisor of $2^{2k} - 1$ and such an element can not exist in $\clifford_{k-1}$. Hence, if we want to implement $\overline{U}$ on the logical qubits of a $[[n,k,d]]$ stabilizer code, by Lemma 1 we need to implement a physical gadget having the same order $p$ which must belong in $\clifford_k \backslash \clifford_{k-1}$ and must be implemented using  a $k$-fold transversal gadget. Hence, Theorem 1 is proved. \qed

\subsection{Proof of Theorem 2}

We now prove Theorem 2 stated in Section \ref{main}, which constrains gates implemented by code automorphisms, including any combination of single-qubit Clifford gates and permutations of the qubits. We show that in particular there is a logical gate, the Bell Gate in  $\clifford_2$, which can't be implemented using code automorphisms and hence, there doesn't exist any stabilizer code such that the full Clifford group on multiple logical qubits can be implemented using code automorphisms.

The Bell gate is Clifford gate that yields a matrix transformation that conjugates any real matrix in $\mathrm{SU}(4)$ to a tensor product of two single qubit gates \cite{hill1997entanglement, vatan2004optimal, kubischta2025classification}. In other words, Bell gate conjugates the subgroup $\mathrm{SO}(4)$ to the subgroup $\mathrm{SU}(2) \otimes \mathrm{SU}(2)$. The matrix transformation is given by, 
\begin{equation}
    \mathrm{BELL} = \frac{e^{\frac{3\pi i}{4}}}{\sqrt{2}}\begin{bmatrix}
        1 & i & 0 & 0 \\
        0 & 0 & i & 1 \\
        0 & 0 & i & -1 \\
        1 & -i & 0 & 0 
    \end{bmatrix}.
\end{equation}

The projective Bell Gate or the equivalent classical symplectic matrix of the Bell gate is given by,
\begin{equation}
    M_\mathrm{BELL} = \begin{bmatrix}
        1 & 1 & 0 & 1 \\
        0 & 1 & 0 & 1 \\
        1 & 0 & 1 & 0 \\
        1 & 1 & 1 & 0 
    \end{bmatrix}.
\end{equation}

The Bell gate belongs to $\clifford_k \: \forall \: k \geq 2$ and thus any code which has the full logical Clifford group on multiple qubits must have the logical Bell gate as well. Since the Bell gate is a Clifford gate, its conjugation action on the Pauli basis is given by, 
\begin{equation}
\begin{split}
        &X \otimes I \Rightarrow  Y\otimes Z \quad\quad Z\otimes I \Rightarrow Z \otimes Z \\
        &I \otimes X \Rightarrow X \otimes Y \quad\quad  I\otimes Z  \Rightarrow X \otimes X.
\end{split}
\end{equation}

The Bell gate (and any power of the Bell gate not divisible by 5) has two important properties which would be used during the proof.

\begin{enumerate}
    \item The Projective Bell Gate is an order 5 gate, i.e., $\mathrm{BELL}^5 = \id_2$.
    \item The Bell gate can't preserve a maximal abelian subgroup of $\mathcal{P}_2$ i.e. for every maximal abelian subgroup $G \subseteq \mathcal{P}_2$ there exists $g \in G$ such that $\{\mathrm{BELL}^\dag\: g \:\mathrm{BELL}, g\} = 0$.
\end{enumerate}

Let us now assume that there exists a stabilizer code for which the logical Bell gate can be implemented using a code automorphism and let's call this automorphism $U'$. According to Lemma 1, $U'$ must have an order which is a multiple of 5. Then without loss of generality let us assume,
\begin{equation}
    \ord(U') = 5^m \times q
\end{equation}
where, $m,q \in \mathbb{Z}$ and $q$ is not divisible by 5. Since the code automorphisms form a group, $U = (U')^q$ will also be a code automorphism which will have exactly order $5^m$ and will implement the logical gate $\overline{\mathrm{BELL}}^{q}$. Note that the logical gate still satisfies both the properties listed above since $q$ is not divisible by 5.

We now show that any automorphism that has an order $5^m$ must be \textbf{$\mathbf{5^m}$ local automorphism}. A $5^m$ local automorphism is defined as a permutation of physical qubits that contain only cycles of order $5^m$ followed by a transversal gadget that acts trivially on the non-permuted qubits (but could be non-trivial on the permuted qubits).

\textbf{Lemma 5. [Locality Lemma]} Any code automorphism that has an order which a power of a prime, $p^m$ with $p > 3$, must be a $p^m$ local automorphism.

\textbf{Proof.} Code automorphism consist of a permutation of the physical qubits followed by a transversal gadget. Since transversal gadgets consist of single qubit Cliffords acting individually on each physical qubit, they can only have order $=1,2, 3 \:\text{or}\: 6$. In order for the code automorphism to have an order $p^m$ with $p > 3$, the permutation must have cycles of order $p^m$.

In addition, if the transversal gadget part of the automorphism acts non-trivially on a non-permuted qubit, then applying the automorphism $p^m$ times would still result in a non-trivial action on that qubit, because the order of the single-qubit Clifford gate on that qubit does not divide $p^m$. The order of the automorphism would then be a multiple of $p^m$. Hence, to have an automorphism of order exactly $p^m$, it must act trivially on the non-permuted qubits. \qed

We have now shown that any stabilizer code that has the logical BELL gate implemented using code automorphisms must have a $5^m$ local automorphism that implements the logical $\mathrm{BELL}^q$ gate for some integers $m,q$ where $q$ is not divisible by 5.

\textbf{Lemma 6. [Equivalence Lemma]} For any $p^m$ local automorphism $U$ there exists a transversal gadget $V$ such that $V^\dag U V$ is just a qubit permutation.

\textbf{Proof.} Without loss of generality let 
\begin{equation}
    U = \left(\underset{1 \leq i \leq p^m}{\otimes} U_i \right)* (1,2,\cdots, p^m),
\end{equation}
where $(1,2, \cdots, p^m)$ is a permutation that cyclically permutes the first $p^m$ physical qubits. Since $U$ has order $p^m$, 
\begin{equation}
    \prod_{i=1}^{p^m} U_i = \id.
\end{equation}

Now, 
\begin{equation*}
\begin{split}
    V^\dag U V = &\Big( \underset{1 \leq i \leq p^m}{\otimes} V_i^\dag  \Big) \Big( \underset{1 \leq i \leq p^m}{\otimes} U_i * (1,2,\cdots, p^m)  \Big) \\
    &\Big( \underset{1 \leq i \leq p^m}{\otimes} V_i  \Big) \\
    =  &\Big( \underset{1 \leq i \leq p^m}{\otimes} V_i^\dag U_i V_{i+1 \: \mathrm{mod}\: p^m}\Big) * (1,2,\cdots, p^m) .
\end{split}
\end{equation*}

So we can choose $V_1 = \id$ and $V_i =  \big( \prod_{j<i} U_j \big)^\dag$ such that 
\begin{equation}
    V^\dag U V = \underset{1 \leq i \leq p^m}{\otimes} \id \: * (1,2,\cdots, p^m). 
\end{equation}

Hence, proved. If the automorphism has multiple permutation cycles of size $p^m$ then the above procedure can be carried out independently for each cycle of permuted qubits since they must be disjoint. \qed

Given Lemma 6, we can construct an equivalent stabilizer code to the one with the logical Bell gate by choosing its stabilizer group to be $V^\dag \: \mathcal{S}\: V $ where $\mathcal{S}$ is the stabilizer group of the original code. Then for the equivalent code, 
\begin{equation}
    \big(V^\dag U V \big)^\dag \big(V^\dag S V \big) \big(V^\dag U V \big) \in V^\dag \: \mathcal{S}\: V \quad \forall \: S \in \mathcal{S}.
\end{equation}

Thus, $V^\dag U V$ is an automorphism for the new code. Moreover, the logical gate implemented by $V^\dag U V$ for the new code must satisfy both the properties of the Bell gate. In addition, $V^\dag U V$ is just a permutation of physical qubits. In the next Lemma we show that a physical gadget which is just a permutation can never satisfy the second property.

\textbf{Lemma 7. [Permutation Lemma]} Any permutation of the physical qubits must implement a logical gate that preserves the abelian subgroup $\{\overline{Z_i}\}$.

\textbf{Proof.} It was shown in \cite{gottesman1997stabilizer}  [Chapter 4, Equation 4.3] that all stabilizer codes in their binary symplectic form can be reduced to a standard form by performing Gaussian elimination.

\begin{figure}[!h]
    \centering
    \includegraphics[width=\linewidth]{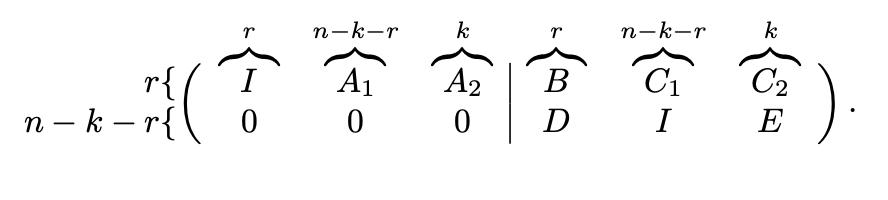}
    \label{stdform}
\end{figure}
\vspace{-1 cm}
For a code in the standard form we can choose a set of logical Pauli representatives, 
\begin{equation}
    L = \begin{bmatrix}
        L_x \\ L_z 
    \end{bmatrix} =  \left[
 \begin{array}{ccc|ccc}
0 & E^T & I & E^TC_1^T + C_2^T & 0 & 0 \\
0 & 0   & 0   & A_2^T & 0 & I
\end{array}
\right].
\end{equation}

The important thing to note is that in the standard form, the logical Z operators are completely made up of physical Z terms.

Now let us consider a permutation of physical qubits $U$. Since $\overline{Z}_i \:\forall\: i$ is completely made up of physical Z terms, $U\overline{Z_i}U^\dag$ must also be completely made up of physical Z terms. As a result $[\overline{Z_i}, U\overline{Z_i}U^\dag] = 0 \: \forall\: i$. \qed

\textbf{Corollary.} There exists no stabilizer code that admits the full Clifford group on the logical qubits using qubit permutations. This is true even for codes with a single logical qubit whereas for code automorphisms the result only holds for codes with multiple logical qubits.

Hence, according to Lemma 7 all permutation gadgets preserve a maximal abelian subgroup ($\{\overline{Z_i}\}$ given one particular choice of logical operators). Since the BELL gate can never preserve a maximal abelian subgroup, permutation gadgets can never implement the logical BELL gate. Following the train of thought backwards, we prove that no stabilizer code can have a code automorphism which implements the logical Bell gate, therefore the full logical Clifford group on multiple qubits. Hence, proved. \qed

\section{Discussion} \label{dis}

In this paper we present and prove two theorems that impose constraints on existing fault-tolerant Clifford architectures such as fold-transversal gadgets and code automorphisms. Firstly, we show that no stabilizer code has a transversal Clifford implementation of the full Clifford group on multiple logical qubits or a fold-transversal Clifford implementation of the full Clifford group on more than two logical qubits. We also study code automorphisms, which are physical gadgets made up of transversal Clifford gates followed by a permutation of physical qubits, and show that there exists no stabilizer code that admits the full Clifford group on multiple logical qubits implemented using code automorphisms. We then introduce the notion of $k$-fold transversal gadgets and show that, in order to implement the full Clifford group on $k$ logical qubits, we need at least $k$-fold transversal gadgets at the physical level.

A $k$-fold transversal gadget can, in the worst case, propagate a single-qubit error onto $k$ physical qubits. Stabilizer codes encoding multiple logical qubits within a single code block are desirable not only to reduce qubit overheads but also because they provide advantages such as more flexible code-switching and gauge-fixing frameworks. It is also desirable to perform the full Clifford group with a simple implementation, such as a $k$-fold transversal implementation for small $k$. However, here we show that finding stabilizer codes with both of these desired properties is not possible within the fault-tolerant frameworks considered in this paper. Thus, to implement the full Clifford group on multiple logical qubits, we need more complex fault-tolerant constructions.

An important feature of our result is that each class of gadget constructions we consider forms a group, in the sense that compositions of such gadgets do not lead to increased fault propagation. Consequently, any subgroup of logical operations realized via code automorphisms or $k$-fold transversal gadgets can be implemented in a depth that depends on $k$ but is constant in the number of physical qubits $n$.
In contrast to our result, adding power beyond $k$-transversal gates can result in more powerful logical gate sets. However, doing so requires non-constant circuit depth and additional fault-tolerant circuitry (such as extra error-correction steps, which typically require significant space-time volume). Two recent works provide examples:
\begin{enumerate}
\item In \cite{tansuwannont2026constructionlogicalcliffordgroup}, the author constructs families of codes with an arbitrary number of logical qubits for which the generators of the logical Clifford group can be implemented using fold-transversal gadgets. However, these constructions rely on fold-transversal operations defined with respect to different partitions. As a result, the corresponding physical gadgets do not form a group, and intermediate error-correction steps are required when composing them. This leads to implementations of logical Clifford operations that are not constant depth. Thus, this approach circumvents the no-go theorem by allowing increased circuit depth.
\item In \cite{malcolm2025computing}, the authors present an efficient scheme for implementing the full logical Clifford group in quantum LDPC (QLDPC) codes. Their approach uses subsystem codes and syndrome-extraction measurements to realize certain Clifford operations. These ingredients fall outside the framework considered here, but they suggest alternative avenues for circumventing the no-go theorem.
\end{enumerate}
Moreover, the theorems established in this work also have important implications for the addressability of stabilizer codes encoding multiple logical qubits. In particular, since addressable logical H, S, and CNOT gates acting on all logical qubits generate the full logical Clifford group, our results imply that no stabilizer code can admit a fully addressable implementation of these gates on all logical qubits using only Clifford transversal constructions or code automorphisms.
\subsection*{Future Directions}

In this work, we have limited our attention only to physical gadgets which belong to the Clifford group, in order to perform logical actions belonging to the Clifford group. It remains an interesting open question whether non-Clifford transversal or fold transversal gates at the physical level could be used as part of a construction of the full logical Clifford group. In addition, we could also investigate the applicability of these theorems to qudit stabilizer codes. As qudit fault tolerance has been explored less than qubit fault tolerance, establishing the existence (or absence) of analogous no-go constraints would be an important contribution to the field. Another possible direction could be to search for stabilizer codes which have certain geometrical advantages that allow for a favorable implementation of $k$-fold transversal gadgets. In such codes, even though $k$- fold transversal gadgets spread errors on to $k$ physical qubits, they only result in high weight errors that correspond to a unique syndromes. These codes are then able to detect and correct such errors resulting from the implementation of the $k$- fold transversal gadget, despite having a smaller distance. A possible example could be to consider $k$ blocks of the $[[7,1,3]]$ code as a single block encoding $k$ logical qubits.

\section{Acknowledgment}

A.C would like to thank Xiaozhen Fu, Victor V Albert, Jin Ming Koh, Shayan Majidy, Alexander Frei and Zachary Mann for their helpful discussions. D.G is
partially supported by the National Science Foundation (RQS QLCI grant OMA-2120575).

\appendix
\onecolumngrid % Switches the layout to single-column
\section{}\label{app1}

Here we illustrate Lemma 3 by explicitly constructing $V,W \in \sym(4,2)$ which correspond to Clifford operators $\in \clifford_2$ acting on 2 qubits. According to the Lemma, $\ord(V) = 2^k - 1= 3$ and $\ord(W) = 2^k + 1= 5$.

\textbf{(a.)} We first choose a primitive polynomial of degree 2,
\begin{equation}
    p(x) = x^2 + x + 1.
\end{equation}

We then construct the field extension $\mathbb{F}_4 \cong \mathbb{F}_2[x]/(p(x))$. Then the companion matrix of $p(x)$ is given by
\begin{equation}
    C_p = \begin{bmatrix}
        0 & 1 \\
        1 & 1
    \end{bmatrix}.
\end{equation}

It can be easily verified that $C_p^3 = \id$ hence $\ord(C_p) = 3$. We then construct $V \in \sym(4,2)$ such that 
\begin{equation}
\begin{split}
    V &= \begin{bmatrix}
        C_p & \mathbf{0}_{k \times k} \\
        \mathbf{0}_{k \times k} & (C^{-1}_p)^T
        \end{bmatrix} \\
        &= \left[
\begin{array}{cc|cc}
0 & 1 & 0 & 0 \\
1 & 1 & 0 & 0 \\
0 & 0 & 1 & 1 \\
0 & 0 & 1 & 0
\end{array}
\right]
\end{split}
\end{equation}

It can be verified that $\ord(V) = \ord(C_p) = 3$ and it satisfies the symplectic condition i.e. $V^T J V = J$. An interesting thing to note is that in this construction $V$ acts on the $X$ and $Z$ Pauli subspaces independently which means it can be realized by a CNOT circuit.

\textbf{(b.)} We select the primitive polynomial 
\begin{equation}
    p(x) = x^4 + x+ 1,
\end{equation}

and construct $\mathbb{F}_{16} = \mathbb{F}_2[x] / (p(x))$. Let $\alpha$ be the residue class of $x$ and since $p(x)$ is a primitive polynomial each element of $\mathbb{F}_{16}$ can be written $\{0, \alpha^k\}$ for $1 \leq k \leq 15$.

In order to make the bilinear form non-degenerate we define the kernel of the bilinear form as, 
\begin{equation}
    \text{Ker}(B) = \{a \in \mathbb{F}_{16} \: |\: B(a,b) = 0 \: \forall \:  b \in \mathbb{F}_{16} \}.
\end{equation}

In order for the standard form to be non-degenerate it's kernel must contain only the zero element i.e. $\text{Ker}(B) = \{0\}$. Simplifying, we can show that in order for the form to be non-degenerate $(\theta + \theta^4) \neq 0$ which implies $\theta \notin \{0,1,\alpha^5, \alpha^{10}\}$. Hence, we choose $\theta = \alpha$ and the non-degenerate standard bilinear form is given by,
\begin{equation}
    B(a,b) = \operatorname{Tr}_{\mathbb{F}_{16}/\mathbb{F}_2}
        \left( \alpha \cdot (a b^4 - a^4 b)\right),
\end{equation}

We now choose a symplectic basis for $\mathbb{F}_{16}$ such that $B(a,b) = a^T J b$. Let the symplectic basis be $\{u_1, u_2, v_1, v_2 \}$ which must satisfy the conditions,
\begin{equation}
    B(u_i, u_j) = 0 \quad B(v_i, v_j) = 0 \quad B(u_i, v_j) = \delta_{ij} \quad \forall \: i,j
\end{equation}

One such symplectic basis can be chosen as $\{1, \alpha^5, \alpha^3, \alpha^4 \}$. Then, $W$ is the linear transformation generated by multiplication by $\alpha^3$.
\begin{equation}
\begin{split}
    W (1) &= \alpha^3 \\
    W (\alpha^5) &= \alpha^8 = \alpha^5 + \alpha^4 \\
    W (\alpha^3) &= \alpha^6 = 1 + \alpha^5 + \alpha^3 + \alpha^4 \\
    W (\alpha^4) &= \alpha^7 = \alpha^3 + \alpha^4 
\end{split}
\end{equation}

Hence, 
\begin{equation}
    W = \left[
\begin{array}{cc|cc}
0 & 0 & 1 & 0 \\
0 & 1 & 1 & 0 \\
1 & 0 & 1 & 1 \\
0 & 1 & 1 & 1
\end{array}
\right]
\end{equation}

It can be verified that $\ord(W) = 5$ and it satisfies the symplectic condition $W^T J W = J$. \qed

\end{document}